\documentclass[letterpaper, 10 pt, journal, twoside]{ieeetran}  %

\IEEEoverridecommandlockouts                              %
\usepackage[pdftex]{graphicx} 
\usepackage{amsmath} %
\usepackage{amssymb}  %
\usepackage{cite}
\usepackage{chemarr}
\newcommand{\setm}{\{m_{\bm \alpha}\}}
\newcommand{\sethm}{\{\hat{m}_{\bm \alpha,s}\}}

\title{Bounding Transient Moments \\of Stochastic Chemical Reactions}
\author{Yuta Sakurai and Yutaka Hori%
\thanks{This work was supported in part by JSPS KAKENHI Grant Number JP16H07175 and JP18H01464, and Keio Gijuku Academic Development Funds.}%
\thanks{Y. Sakurai and Y. Hori are with Department of Applied Physics and Physico-Informatics, Keio University. 3-14-1 Hiyoshi, Kohoku-ku, Yokohama, Kanagawa 223-8522, Japan. {\tt\small y.sakurai-5861@keio.jp}, {\tt\small yhori@appi.keio.ac.jp}}%
}
\usepackage{bm}
\def\coloneqq{\mathrel{\mathop:}=}

\usepackage{url}
\usepackage{fancyhdr}

\fancypagestyle{firstpage}{\lfoot{\scriptsize \copyright 2018 IEEE. Personal use of this material is permitted. Permission from IEEE must be obtained for all other uses, in any current or future media, including reprinting/republishing this material for advertising or promotional purposes, creating new collective works, for resale or redistribution to servers or lists, or reuse of any copyrighted component of this work in other works. The final version of record is available at \url{http://dx.doi.org/10.1109/LCSYS.2018.2869639}} \cfoot{}}

\begin{document}

\maketitle
\thispagestyle{firstpage}

  \begin{abstract}
The predictive ability of stochastic chemical reactions is currently
   limited by the lack of closed form solutions to the governing
   chemical master equation. 
   To overcome this limitation, this paper proposes a computational method
   capable of predicting mathematically rigorous upper and lower bounds
   of transient moments for reactions governed by the law of mass action.
   We first derive an equation that transient moments must satisfy based
   on the moment equation. Although this equation is underdetermined, we 
   introduce a set of semidefinite constraints known as moment condition to narrow
   the feasible set of the variables in the equation.
   Using these conditions, we formulate a semidefinite program that efficiently and rigorously computes the bounds of transient moment dynamics.
   The proposed method is demonstrated with illustrative numerical
   examples and is compared with related works to discuss advantages and limitations. 
  \end{abstract}

\section{Introduction}
\par
\smallskip
\IEEEPARstart{C}{hemical} processes in living cells are highly stochastic unlike those in
engineered batch reactors because of the low abundance of reacting
molecules in a cell. The stochastic reactions result in large
cell-to-cell variations of molecular copy numbers and lead to phenotypic
diversity, which is often considered to be beneficial for bet hedging
against perturbations. Thus, in biology, the dynamics of cellular
reactions are often measured as the snapshots of population
distributions rather than a trajectory of a single representative cell
(see \cite{Gardner2000,Nevozhay2012,Wu2013,Hsiao2016,Biggar2001} for example).

\par
\smallskip
The dynamics of stochastic chemical reactions are modeled by a discrete
state Markov process, where the state represents the copy number of
molecules. The evolution of the stochastic process is, thus,
characterized by Kolmogorov forward equation, also known as the chemical
master equation (CME) \cite{Gillespie1992}. Unfortunately, an exact
analytic solution to the CME is not known  except for some simple
reactions due to the fact that the state of the Markov chain is
semi-infinite, {\it i.e.} a set of non-negative integers. Consequently,
the stochastic chemical reactions are currently analyzed by time
consuming sample-path simulations \cite{Gillespie1976} or approximation
based techniques such as the finite state projection
\cite{Munsky2006,Gupta2017}, linear noise approximation
\cite{vanKampen2007} and Langevin equations \cite{Gillespie2000, Gillespie2001, Marquez-Lago2007,
Melykuti2010}. %

\par
\smallskip
Other approaches attempt to directly compute the moments of the
stochastic process based on moment equation, the governing equation of
moment kinetics derived from the CME.
This approach enables a direct characterization of the statistics of stochastic chemical reactions such as the mean and the
covariance of molecular abundance. However, the moment equation
essentially faces the same problem as the CME --- the equation forms an infinite chain of ODEs, and the solution is
analytically intractable for most stochastic reactions of interest.
To deal with this issue, moment closure \cite{Zhao2010,
 Singh2011,Lakatos2015} is widely used to obtain a (truncated) finite order equation by 
 approximately expressing high order moments with low order moments at
 the cost of accuracy. 

 \par
 \smallskip
On the other hand, computing moments with guaranteed precision remains an active research topic. 
In \cite{Gupta2014}, a recursive algorithm was proposed to obtain bounds of moments based on concentration
 inequalities.
 More recently, a semidefinite program (SDP) \cite{Boyd2004} was formulated to compute
 guaranteed upper/lower bounds of steady state moments based on the moment equation \cite{SakuraiCDC2017, Ghusinga2017, Kuntz2017, DowdyConf2017,
 Sakurai2018, Dowdy2018}. 
These works also used the truncated moment equation used in the moment
 closure, but they compensated for the truncated moments based on a relaxation that a moment matrix, a matrix defined by a product of monomial vectors, is positive semidefinite.
 The use of the semidefinite relaxation was motivated by its close connection with a so-called {\it moment condition}, a necessary and sufficient condition for a given sequence of real numbers to be moments of some non-negative measure (probability
 distribution) under some assumptions \cite{Landau1987}.
Although complete understanding the underlying mechanism requires
 further study, the previous works demonstrated that this approach could
 give surprisingly tight bounds of  steady state moments only with a small number of moments \cite{SakuraiCDC2017, Ghusinga2017, Kuntz2017, DowdyConf2017, Sakurai2018, Dowdy2018}.

\par
\smallskip
Building upon the idea for the steady state moment computation
\cite{SakuraiCDC2017, Sakurai2018}, this paper presents a semidefinite
program capable of computing the upper and lower bounds of transient, or
dynamic, moments for stochastic reactions consisting of elementary reactions.
An obvious requirement for this extension is the introduction of new variables and constraints for the
transient moments. For this purpose, we introduce ``temporal moments'', the moments of state
variables and time. Conceptually, this means that we regard the time 
variable $t$ as part of random variables and attempt to compute the moments of 
a measure supported on the state space of the Markov chain and a real number. 
This leads to a new equality constraint that replaces the steady state 
moment equation used in the previous study \cite{SakuraiCDC2017,
Sakurai2018} and introduces additional semidefinite conditions that
correspond to moment conditions.
Consequently, we obtain a semidefinite program for computing 
the transient statistics of molecular abundance.

\par
\smallskip
It should be noted that, recently, a similar approach was developed in 
parallel by Dowdy and Barton \cite{Dowdy2018arXiv}.
A main difference from the proposed method is that a moment generating function of
$t$ is considered instead of a moment itself to constrain the values of 
transient moments. This results in different semidefinite conditions from
the proposed method, and the computed bounds are indeed different. 
Here we also discuss how these different formulations affect the tightness of
the bounds by comparing the results for multiple reaction examples.

\par
\smallskip
This paper is organized as follows. In Section II.A, we introduce the
moment equation. Then, in Section II.B, we define the temporal moment and
formulate the optimization problem. %
Illustrative numerical examples are provided in Section III, and the
results are compared between the proposed approach and the one in \cite{Dowdy2018arXiv}.
Finally, Section IV concludes this paper.

\par
\smallskip
The following notations are used in this paper. 
$\mathbb{N}\coloneqq\{1,2,3,\cdots\}$. $\mathbb{N}_0\coloneqq \mathbb{N} \cup \{0\}$.
$\mathbb{Z}$ is a set of all integers. %
$\mathbb{R}_+\coloneqq\{x \in \mathbb{R}~|~x \ge 0\}$.
$\mathbb{R}[{\bm x}]$ is the set of all polynomials with real coefficients. 
${\rm deg}(p({\bm x}))\coloneqq\sum_{j=1}^{n}p_j$ is the degree of a monomial $p({\bm x})=\prod_{j=1}^{n}x_j^{p_j}$.
\section{Computation of moment dynamics of stochastic chemical reactions}
\subsection{Moment dynamics of stochastic chemical reactions}
In this section, we introduce a general mathematical model of stochastic
chemical reactions and review an ordinary differential equation (ODE)
model of moment dynamics. Consider a chemical reaction system that
consists of $n\in \mathbb{N}$ species of molecules and $r$ types of
reactions. The copy numbers of the $n$ molecules are denoted by
$\bm{x}\coloneqq[x_1,x_2,\cdots,x_n]^\mathrm{T}\in \mathbb{K}$, where
$\mathbb{K} \subseteq \mathbb{N}_0^n$ represents a set of all possible
combinations of copy numbers. 
The copy numbers $\bm{x}$ specify the state of the reaction system and
fluctuate in time due to stochastic chemical reactions. The stochastic
fluctuation of the copy numbers $\bm{x}$ can be modeled by a Markov
process. Specifically, we define $P_{\bm{x}}(t)$ as the probability that
there are $\bm{x}$ molecules at time $t$. Then the dynamics of
$P_{\bm{x}}(t)$ follows the following Chemical~Master~Equation~(CME) \cite{Gillespie1992},
\begin{align}
\frac{dP_{\bm{x}}(t)}{dt}=\sum_{i=1}^r\left\{w_i(\bm{x}-\bm{s}_i)P_{\bm{x}-\bm{s}_i}(t)-w_i(\bm{x})P_{\bm{x}}(t)\right\}\label{eq:CME},
\end{align}
where $w_i(\bm{x})$ is the propensity function (reaction rate) of the
$i$-th reaction ($i=1,2,\cdots,r$), and
$\bm{s}_i=[s_{i1},s_{i2},\cdots,s_{in}]^\mathrm{T}~\in\mathbb{Z}^n$ is
the stoichiometry of the $i$-th reaction. We assume that all reactions
are elementary. That is, each reaction is either a unimolecular or a
bimolecular reaction, and the propensity function $w_i(\cdot)$ is a
polynomial of $x_j~(j=1,2,\cdots,n)$ \cite{Denisov2003}.

\par
\smallskip
To derive an ODE model of moment dynamics based on the CME (\ref{eq:CME}), we define a raw moment of a probability distribution $P_{\bm{x}}(t)$ by
\begin{align}
m_{\bm{\alpha}}(t)\coloneqq\mathbb{E}_t\left[\prod_{j=1}^nx_j^{\alpha_j}\right]=\sum_{\bm{x}
 \in \mathbb{K}}\prod_{j=1}^nx_j^{\alpha_j}P_{\bm{x}}(t),\label{eq:defmoment}
\end{align}
where
$\bm{\alpha}\coloneqq[\alpha_1,\alpha_2,\cdots,\alpha_n]\in\mathbb{N}_0^n$. 
We then multiply $\prod_{j=1}^{n} x_j^{\alpha_j}$ to both sides of the CME
(\ref{eq:CME}) and take sum over $\mathbb{K}$ to obtain
an ODE of moments known as moment equation
\begin{align}
\frac{d}{dt}\bm{m}(t)=A\bm{m}(t)+B\bm{u}(t), \label{eq:StateSpaceRealization}
\end{align}
where 
 $A$ and $B$ are constant matrices whose entries are linear combinations
 of the rate constants of the propensity functions $w_i(\cdot)$, and 
$\bm{m}(t)$ and $\bm{u}(t)$ are vectors of raw moments up to the
$\mu$-th order and those of the $\mu+1$-th order, respectively (see
Section 2.2 and S.3 of \cite{Sakurai2018} for derivation). 
The moment equation (\ref{eq:StateSpaceRealization}) implies that the
raw moments of the $\mu+1$-th order, {\it i.e.}, the entries of
$\bm{u}(t)$, are required for computing the moments up to the $\mu$-th
order, {\it i.e.}, the entries of $\bm{m}(t)$. Thus, we need to estimate
$\bm{u}(t)$ to obtain the solution $\bm{m}(t)$ of the moment equation.

\par
\smallskip
In the next section, we present an approach to solving the equation (\ref{eq:StateSpaceRealization}) without explicitly computing $\bm{u}(t)$. The proposed approach utilizes a so-called moment condition to find the lower and upper bounds of the moments $\bm{m}(t)$ of molecular copy numbers $\bm{x}$ at time $t$. This allows us to rigorously bound the transient statistics of the copy numbers $\bm{x}$ over time.

\subsection{Semidefinite programming for transient moment analysis}
In this section, we present a mathematical optimization problem for rigorously bounding the transient moments. Our derivation is based on the recently proposed method for computing the steady state moments \cite{SakuraiCDC2017, Ghusinga2017, Kuntz2017, DowdyConf2017, Sakurai2018, Dowdy2018}.

\par
\smallskip
Let $\hat{\bm{m}}_s(T_1,T_2)$ and $\hat{\bm{u}}_s(T_1,T_2)$ be defined by 
\begin{align}
\hat{\bm{m}}_s(T_1,T_2)&\coloneqq\int_{T_1}^{T_2}t^s \bm{m}(t)dt\label{eq:DefMhatVector},\\
\hat{\bm{u}}_s(T_1,T_2)&\coloneqq\int_{T_1}^{T_2}t^s
 \bm{u}(t)dt, \label{eq:DefUhatVector}
\end{align}
where $s \in \mathbb{N}_0$. 
We derive an equation of these vectors by multiplying $t^s$ to both sides of the moment equation (\ref{eq:StateSpaceRealization}) and taking the integral of time $t$ as 
\begin{align}
 \int_{T_1}^{T_2}t^s\frac{d}{dt}\bm{m}(t)dt=A\int_{T_1}^{T_2}t^s\bm{m}(t)dt+B\int_{T_1}^{T_2}t^s\bm{u}(t)dt. \notag
\end{align}
Using integration by parts, this equation can be calculated
as 
\begin{align}
 &{T_2}^s\bm{m}(T_2)-{T_1}^s\bm{m}(T_1)-s\hat{\bm{m}}_{s-1}(T_1,T_2)\notag \\
 &=A\hat{\bm{m}}_s(T_1,T_2)+B\hat{\bm{u}}_s(T_1,T_2),\label{eq:TimeMomentEquation}
\end{align}
where we define $0^0=1$ in the case of $T_i = 0~(i=1,2)$ and $s=0$.

\par
\smallskip
The transient values of raw moments at time $T_1$ and $T_2$ could be obtained if we could solve the linear equation (\ref{eq:TimeMomentEquation}). In particular, if we have a priori knowledge of the moments at the initial time, say $T_1 = 0$, the transient moment computation reduces to finding $\bm{m}(T_2)$ for the given initial moment $\bm{m}(T_1)$. In general, however, the equation (\ref{eq:TimeMomentEquation}) is underdetermined, and the solution is given only as a certain linear subspace.

\par
\smallskip
Thus, we need more conditions to further specify the solution space of
the moments $\bm{m}(T_2)$, $\bm{u}(T_2)$, $\hat{\bm{m}}_s(T_1,T_2)$ and $\hat{\bm{u}}_s(T_1,T_2)$.
Here, we use the fact that the entries of 
these vectors must be moments of some non-negative measure. It should be
noted that $\hat{\bm{m}}_s(T_1,T_2)$ and $\hat{\bm{u}}_s(T_1,T_2)$ can
be viewed as moments of a non-negative measure defined on $\mathbb{K} \times [T_1, T_2]$.

\par
\smallskip
To this end, we start by introducing conditions for $\bm{m}(T_2)$ and
$\bm{u}(T_2)$ to be moments. 
Let
$\bm{X}\coloneqq[(\bm{x}^0)^{\mathrm{T}},(\bm{x}^1)^{\mathrm{T}},\cdots,(\bm{x}^{\gamma_1})^{\mathrm{T}}]^\mathrm{T}$
with $\bm{x}^p$ being a vector of monomial bases satisfying ${\rm
deg}(\prod_jx_j^{p_j})=p$.
It then follows that the entries of the matrix
\begin{align}
 H_0(\{m_{\bm \alpha}\})
 &\coloneqq\mathbb{E}_t[\bm{X}\bm{X}^\mathrm{T}]=\sum_{\bm{x} \in
 \mathbb{K}}\bm{X}\bm{X}^\mathrm{T}P_{\bm{x}}(t)
 \label{H0-def}
\end{align}
consist of the moments $m_{\bm{\alpha}}(t)$, which are the entries of
${\bm m}$ and ${\bm u}$. Moreover, $H_0(\{m_{\bm \alpha}\}) \succeq O$
holds due to its definition (\ref{H0-def}). 
Thus, $H_0(\{m_{\bm \alpha}\})$ constitutes a linear matrix inequality (LMI)
condition that the entries of ${\bm m}$ and ${\bm u}$ must satisfy.

\par
\smallskip
The moment values can further be constrained by using the fact that the moments are defined
for the measure $P_{{\bm x}}(t)$ on $\mathbb{K}$.
Let $\overline{\mathbb{K}}$ denote a semi-algebraic set specified by 
real polynomials $g_k({\bm x})$, {\it i.e.}, 
 \begin{align}
\overline{\mathbb{K}} := \{{\bm  x}\in \mathbb{R}^n~|~g_k({\bm x}) \ge
  0, g_k \in \mathbb{R}[{\bm x}]~(k=1,2,\cdots,\ell)\},
  \notag
 \end{align}
 and satisfying $\mathbb{K} \subseteq \overline{\mathbb{K}} (\subseteq
 \mathbb{R}^n)$.
We can then obtain a condition $H(\{m_{\bm \alpha}\}, g_k) \succeq O$, 
which ${\bm m}$ and ${\bm u}$ must satisfy, where 
  \begin{align}
H(\{m_{\bm \alpha}\}, 
   g_k)&\coloneqq \mathbb{E}_t[g_k(\bm{x})\bm{X}\bm{X}^\mathrm{T}]
   \notag \\
   &=\sum_{\bm{x}
   \in \mathbb{K}}g_k(\bm{x})\bm{X}\bm{X}^\mathrm{T}P_{\bm{x}}(t). \label{H-def}
  \end{align}

 \noindent
 {\bf Example.}
When the state space of the Markov chain in (\ref{eq:CME})
 is $n=1$ dimension and non-negative integers, {\it i.e.,} $\mathbb{K} =
 \mathbb{N}_0$, we can define $\overline{\mathbb{K}} = \mathbb{R}_+$
 with $g_1(x) = x$. Then, the moments of the probability distribution $P_{x}(t)$ must satisfy
 \begin{align}
  (\ref{H0-def}) =
  \begin{bmatrix}
   m_0(t) & m_1(t)\\
   m_1(t) & m_2(t)
  \end{bmatrix}
 \! \succeq \!O,~
  (\ref{H-def})=
\begin{bmatrix}
 m_1(t)  & m_2(t) \\
 m_2(t) & m_3(t)
\end{bmatrix}
  \!  \succeq O\!
  \notag
 \end{align}
 for each $t$, where $\gamma_1 = 1$ in this example.
Since principal minors of a positive
semidefinite matrix are non-negative, 
the second inequality implies non-negativity of the mean 
$m_1(t) \ge 0$, which indeed constrains the moment value.

\par
\smallskip
In summary, we have the following proposition.

\medskip
\noindent
{\bf Proposition 1.}
Consider $m_{\bm{\alpha}}(t)~(\bm{\alpha}\in\mathbb{N}_0^{n})$ defined
by (\ref{eq:defmoment}), which are the moments of the probability measure $P_{\bm{x}}(t)$ defined on $\mathbb{K}$. 
Then, the following LMIs hold. %
\begin{align}
 H_0(\setm)&\succeq O\label{eq:H0},\\
 H(\setm, g_k)&\succeq O~~~(k=1,2,\cdots,\ell)\label{eq:H},
\end{align}
 where $H_0(\setm)$ and $H(\setm, g_k)$ are
 defined in (\ref{H0-def}) and (\ref{H-def}), respectively.

 \par
 \medskip
These LMIs serve as additional convex constraints to restrict the possible values of the moments
$\bm{m}(\cdot)$ and ${\bm u}(\cdot)$ in the linear equation (\ref{eq:TimeMomentEquation}).
Thus, combining the LMIs (\ref{eq:H0}) and (\ref{eq:H}) with (\ref{eq:TimeMomentEquation}), 
we can formulate a semidefinite program (SDP) that computes upper and
 lower bounds of moment values, which will be seen in detail at
 the end of this section.
 In general, we can obtain progressively tighter bounds 
 by increasing the order of the moments $\mu$ of the vector ${\bm m}$ 
 and the associated LMI conditions, {\it i.e.,}  $\gamma_1$ in
 (\ref{H0-def}) and (\ref{H-def}) since it increases the number of
 equalities (\ref{eq:TimeMomentEquation}) and associated inequality conditions.
 
 \par
 \smallskip
\noindent
{\bf Remark 1.~}
The LMIs (\ref{eq:H0}) and (\ref{eq:H}) can be viewed a necessary condition for a given sequence
 $\setm$ to be moments of some positive measure supported on $\overline{\mathbb{K}}$. 
 It is known that, in some cases, the LMIs %
 are also sufficient.
 In fact, the semidefinite conditions are known as moment condition and have been studied for more than a century (see
 \cite{Landau1987} for example).
 For instance,
for $n=1$ (univariate moments), the LMIs (\ref{eq:H0}) and (\ref{eq:H}) with
 $g_1(x)=x$ and $\gamma_1 \rightarrow \infty$ become a necessary and sufficient condition for the existence of a
 positive measure supported on $\mathbb{R}_+$ \cite{Landau1987}.
 A similar LMI based sufficient condition is known 
 when $n \ge 1$ and $\overline{\mathbb{K}}$ is compact (K-moment
 condition) \cite{Schmudgen1991}. 
 The use of the moment conditions %
 in our work is motivated by these necessary and sufficient conditions, though
 there remain many open problems regarding the sufficiency.

\par
\medskip
Using the same approach, we derive conditions that constrain possible
values of $\hat{\bm m}_s(T_1, T_2)$ and $\hat{\bm u}_s(T_1, T_2)$. 
Each entry of these vectors is represented by 
\begin{align}
\hat{m}_{\bm{\alpha},s}(T_1,T_2)&=\int_{T_1}^{T_2}t^s \sum_{\bm{x} \in \mathbb{K}}
 \prod_{j=1}^nx_j^{\alpha_j}P_{\bm{x}}(t)dt.\label{eq:DefMhat} 
\end{align}
This can be viewed as a moment of a measure defined on $\mathbb{K} \times [T_1, T_2]$.
Thus, it is possible to derive a similar condition to Proposition 1.
We define a vector $\hat{\bm{X}}$ by
$\hat{\bm{X}}\coloneqq[\bm{X}^{\mathrm{T}}t^0,\bm{X}^{\mathrm{T}}t^1,\cdots,\bm{X}^{\mathrm{T}}t^{\gamma_2}]^\mathrm{T},$
and real polynomials $\hat{g}_k({\bm x},t)~(k=1,2,\cdots, \ell')$ that
specify a semi-algebraic set %
\begin{align}
\overline{\mathbb{K}} \times [T_1, T_2] = \{& ({\bm x}, t) \in
 \mathbb{R}^n \times \mathbb{R} ~|~\hat{g}_k({\bm x}, t) \ge
 0,\notag \\&\hat{g}_k \in \mathbb{R}[{\bm x, t}]~(k=1,2,\cdots,\ell') \}. 
\end{align}
Then the following proposition holds.

\medskip
\noindent
{\bf Proposition 2.}~
Consider
$\hat{m}_{\bm{\alpha},s}(T_1,T_2)~(\bm{\alpha}\in\mathbb{N}_0^{n},~s\in\mathbb{N}_0)$
defined by (\ref{eq:DefMhat}).
Let
$\hat{H}_0(\sethm)$ and $\hat{H}(\sethm, \hat{g}_k)$ be
\begin{align}
 \hat{H}_0(\sethm)&\coloneqq\int_{T_1}^{T_2}~\mathbb{E}_t[\hat{\bm{X}}\hat{\bm{X}}^\mathrm{T}]dt\notag
 \\&=\int_{T_1}^{T_2}~\sum_{\bm{x}
 \in \mathbb{K}}\hat{\bm{X}}\hat{\bm{X}}^\mathrm{T}P_{\bm{x}}(t)dt, \label{hH0-def}\\
\hat{H}(\sethm,
 \hat{g}_k)&\coloneqq\int_{T_1}^{T_2}~\mathbb{E}_t[\hat{g}_k(\bm{x},t)\hat{\bm{X}}\hat{\bm{X}}^\mathrm{T}]dt\notag
 \\
 &=\int_{T_1}^{T_2}~\sum_{\bm{x}
 \in
 \mathbb{K}}\hat{g}_k(\bm{x},t)\hat{\bm{X}}\hat{\bm{X}}^\mathrm{T}P_{\bm{x}}(t)dt.
 \label{hH-def}
\end{align}
 Then, the following LMIs hold. %
 \begin{align}
\hat{H}_0(\sethm)&\succeq O\label{eq:Hhat0},\\
\hat{H}(\sethm,\hat{g}_k)&\succeq
  O~~~(k=1,2\cdots,\ell'). \label{eq:Hhat}
 \end{align}

The proof is clear from the definitions (\ref{hH0-def}) and (\ref{hH-def}).
Similar to Proposition 1, the LMIs (\ref{eq:Hhat0}) and (\ref{eq:Hhat})
 become constraints for the values of the moments $\hat{\bm
 m}_s(T_1, T_2)$ and $\hat{\bm u}_s(T_1, T_2)$. 
 In the case of $\overline{\mathbb{K}} = \mathbb{R}_+^n$, the function
 $\hat{g}_k({\bm x}, t)$ can, for example, be defined as
\begin{align}
 \hat{g}_k(\bm{x},t)&=\left\{
 \begin{array}{cl}
x_k&(k=1,2,\cdots,n)\\
t-T_1&(k=n+1)\\
T_2-t&(k=n+2)
 \end{array}\right. .
 \label{hatg-def}
\end{align}

 \medskip
 \noindent
{\bf Remark 2.}~
There can be many possible choices of polynomials $g_k({\bm x})$ and
$\hat{g}_k({\bm x}, t)$ to represent  $\overline{\mathbb{K}}$ and
$\overline{\mathbb{K}} \times [T_1, T_2]$, respectively.
The choice of the polynomials may affect the tightness of the bounds of the
moments, but it is left open to explore what choices give tighter bounds
 in general.

\medskip
\noindent
{\bf Optimization problem}~
Finally, we combine the equation (\ref{eq:TimeMomentEquation}) and the
LMIs (\ref{eq:H0}), (\ref{eq:H}), (\ref{eq:Hhat0}) and (\ref{eq:Hhat}) to formulate an optimization problem that computes the bounds of the statistics of the copy numbers $\bm{x}$.
Consider the stochastic chemical reaction modeled by the equation
(\ref{eq:CME}). The following optimization problem gives the lower bound
of a statistics $f(\bm{m}(t),\bm{u}(t))$ or the upper bound of $-f(\bm{m}(t),\bm{u}(t))$.
 \begin{align}
\begin{array}{rl}
{\rm min} & f(\bm{m}(t),\bm{u}(t))\\
{\rm s.t.} & (\ref{eq:TimeMomentEquation}),~(\ref{eq:H0}),~(\ref{eq:H}),~(\ref{eq:Hhat0})~{\rm and}~(\ref{eq:Hhat})
\end{array}\label{eq:OptiProb}
 \end{align}
The matrices $H_0(\setm)$, $H(\setm,g_k)$, $H_0(\sethm)$ and
$H(\sethm,\hat{g}_k)$ should contain all the moments that appear in the
equation (\ref{eq:TimeMomentEquation}).
Thus, $\gamma_1$ and $\gamma_2$, which determines the size of the vectors
${\bm X}$ and $\hat{{\bm X}}$ should be defined depending on $
\mu$, the highest order of the moments in ${\bm m}$, and $\nu$,
the highest exponent of time in $\hat{\bm{m}}_s(T_1,T_2)$ in the equation
 (\ref{eq:TimeMomentEquation}), respectively (see Appendix for
 definitions). 
 In our optimization, $\mu$ and $\nu$ are tuning parameters that 
 control the tradeoff between the tightness of bounds and computational cost.

\par
\smallskip
The constraints of the optimization problem (\ref{eq:OptiProb}) consist
of a set of linear equalities and semidefinite matrices. Thus, (\ref{eq:OptiProb}) can be formulated as SDP if
$f(\bm{m}(t),\bm{u}(t))$ is linear.
In fact, the computation of many popular statistical values including variance and coefficient of
variations can be converted to the form of SDP even if they are not
linear in raw moments as shown in \cite{Sakurai2018}.

\begin{table}[tb]
\centering
\caption{List of reactions and associated definitions}
 \label{tbl:allwi}
\begin{tabular}{c|ccc}
	\hline  
 		Index & Reaction & Propensity & Stoichiometry $s_i$\\
 		$i$ & & $w_i(x)$ & of ${\rm P}$\\
 		\hline
 		$1$&D $\rightarrow$ P& $k_1D_T$&1\\
 		$2$& P $\rightarrow \phi$ & $k_2x$&-1\\
 		$3$& P+P $\rightarrow$ {\rm P:P}&$k_3x(x-1)$&-2\\
 		\hline
\end{tabular}
\end{table}

  \begin{figure}[tb]
   \centering
  \includegraphics[clip, width=8.5cm]{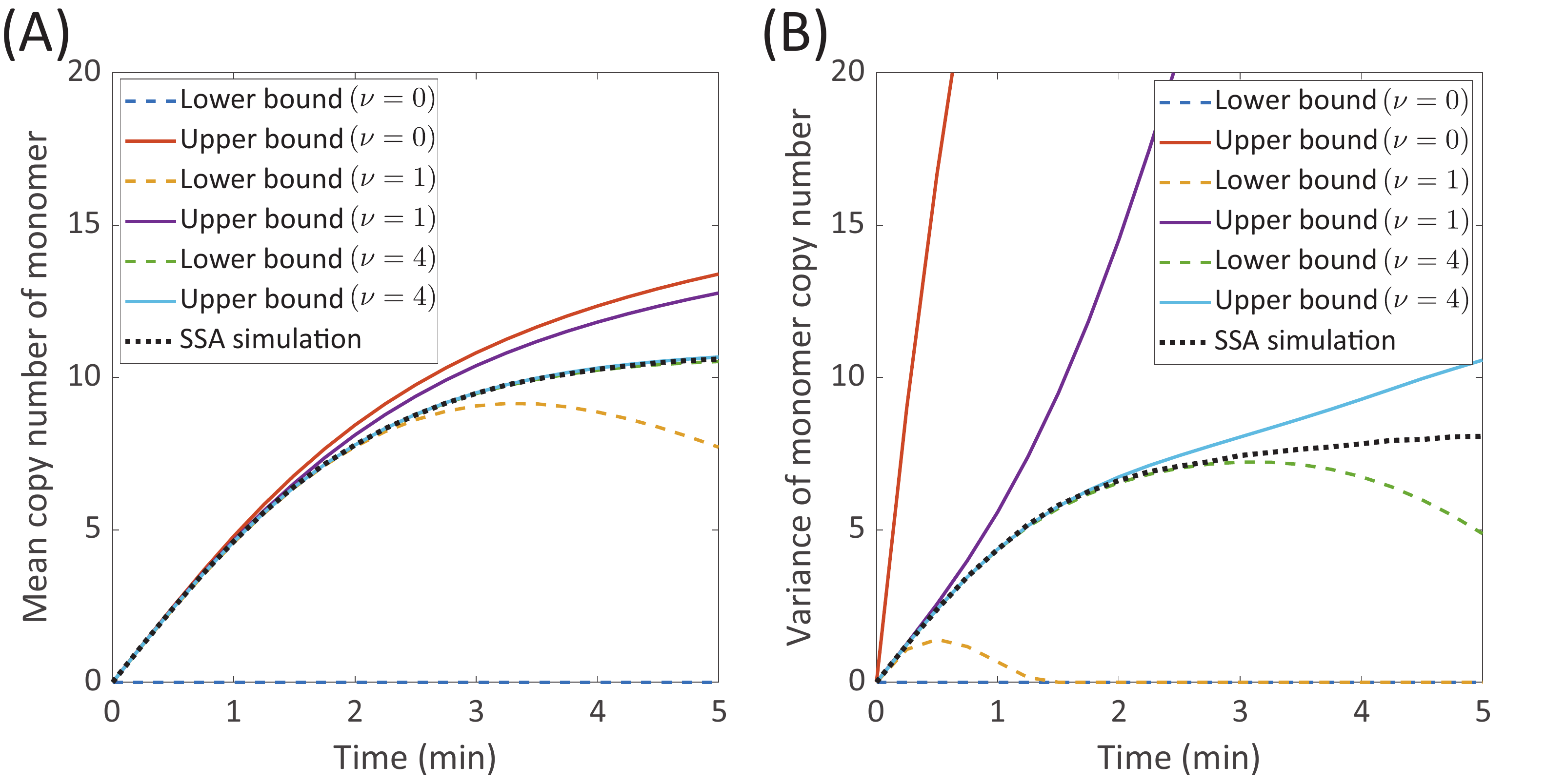}
   \caption{Bounds of the statistics of monomer protein P using the proposed
   approach with  $\mu=7$. The SSA simulation \cite{Gillespie1976} is based on 50,000 sample paths.
   (A)  the mean copy number (B) the variance of the copy number. }
   \label{dimer-fig}
  \end{figure}
\section{Application to stochastic chemical reactions}
  In this section, we first present illustrative numerical examples, and then compare
  the results with a related work that was recently developed in \cite{Dowdy2018arXiv}.

\subsection{Dimerization model with infinite states}
We consider a protein dimerization process that consists of $r=3$
reactions in Table \ref{tbl:allwi}, where ${\rm D}$, ${\rm P}$ and ${\rm P:P}$ represents DNA, monomer
protein and its dimer, respectively, and $x$ is the copy number of the monomer ${\rm P}$. 
Since $x$ can take any non-negative integers, the domain of $x$, or the state space of the Markov chain in
(\ref{eq:CME}), is $\mathbb{K} = \mathbb{N}_0$, which is unbounded. As a
result, the CME (\ref{eq:CME}) becomes an infinite dimensional linear
ODE in terms of $P_{\bm x}(t)$, whose solution is analytically intractable.

\par
\smallskip
In what follows, we analyze the mean and the variance of the monomer
protein ${\rm P}$ at the transient state. 
To this end, we consider a truncated moment equation
(\ref{eq:StateSpaceRealization}) with $\mu=7$, 
where the moment vectors are 
${\bm m}(t) = [m_0(t), m_1(t), \cdots, m_7(t)]^{\rm T}$ and $u(t) = m_8(t)$. 
Note that, by definition (\ref{eq:defmoment}), $m_0(t) = 1$, and the
mean and the variance of the copy number $x$ are $m_1(t)$ and $m_2(t) - m_1^2(t)$, respectively. 
To compute the transient values of the moments, we define the vectors of
moments $\hat{\bm{m}}_s(0,T)$ and $\hat{u}_s(0,T)$ by
(\ref{eq:DefMhatVector}) and (\ref{eq:DefUhatVector}). 
We assume the initial state is $x(0)=0$ for all the cells (samples), that is, $P_0(0)=1$.
Then, we obtain the equality constraints (\ref{eq:TimeMomentEquation}), which can be
represented by 
\begin{align}
 &
 \left[\begin{array}{c}
\bm{m}(T)\\
T\bm{m}(T)
       \end{array}\right]-\left[\begin{array}{c}
I \\
O 
 \end{array}\right]\bm{m}(0) \notag \\
& =\left[\begin{array}{cc}
A & O \\
I & A 
	\end{array}\right]\left[\begin{array}{c}
\hat{\bm{m}}_0(0,T)\\
\hat{\bm{m}}_1(0,T)
\end{array}\right]
+\left[\begin{array}{cc}
\bm{b} & O \\
O & \bm{b}
\end{array}\right]\left[\begin{array}{c}
\hat{u}_0(0,T)\\
\hat{u}_1(0,T)
 \end{array}\right]%
 \notag
\end{align}
where $\bm{m}(0)=[1,0,0,\cdots,0]^{\rm T} \in \mathbb{R}^8$, and the highest exponent of
time is set as $\nu=1$ for an illustration purpose.
This equation is underdetermined as the low order moments
$\bm{m}(T),~\hat{\bm{m}}_0(0,T)$ and $\hat{\bm{m}}_1(0,T)$ are dependent
on the high order moments $\hat{u}_0(0,T)$ and
$\hat{u}_1(0,T)$. Thus, it is impossible to uniquely determine the
solution only from the equality constraint.
Hence, we consider the moment conditions (\ref{eq:H0}), (\ref{eq:H}),
(\ref{eq:Hhat0}) and (\ref{eq:Hhat}) to narrow the solution space.
In this example, we used $g_1(x)=x$ and $\hat{g}_k(x,t)$ shown 
in (\ref{hatg-def}) to represent $\overline{\mathbb{K}} = \mathbb{R}_+
(\supseteq \mathbb{K})$ and $\mathbb{R}_+ \times [T_1, T_2]$, respectively.

\par
\smallskip
Based on this formulation, the bounds of the mean and the variance were 
computed by solving the optimization problem (\ref{eq:OptiProb}) with
MATLAB 2016b and SeDuMi 1.32 solver \cite{Sturm1999}. 
Specifically, we solved the optimization problem for $T = 0.25, 0.50, 0.75,
\cdots, 5.0$.
The parameters were set as $k_1 = 0.1$ min$^{-1}$, $k_2 = {\rm
ln}(2)/20$ ${\rm min}^{-1}$, $k_3 = 0.02$ min$^{-1}$, $D_T = 50$.
To avoid numerical instability, the variables were normalized by
constants (see Implementation Details in Supplementary Material). 
Figure \ref{dimer-fig} (A) and (B) illustrate the bounds of the mean and the variance
of the monomer copy number $x$ for different values of $\nu$, the highest exponent of time in (\ref{eq:TimeMomentEquation}). 
We observe that the upper and lower bounds approach to each other as we 
increase $\nu$. We can also confirm that they are indeed upper/lower
bounds of the statistics by comparing with the sample path simulations
of stochastic simulation algorithm (SSA) \cite{Gillespie1976}.

\par
\smallskip
Regarding computational efforts, it took 0.580 s (CPU time) in 
average to solve a single optimization with a fixed $T$ for $\mu = 7$
and $\nu = 1$ and 4.45 s for $\mu=7$ and $\nu=4$ (see Fig. S1 for more data).
Since the number of decision variables increases combinatorially with
the number of chemical species, $n$, the proposed approach is currently
limited in terms of the size of the reaction networks.

    \begin{figure}[tb]
   \centering
  \includegraphics[clip, width=8.5cm]{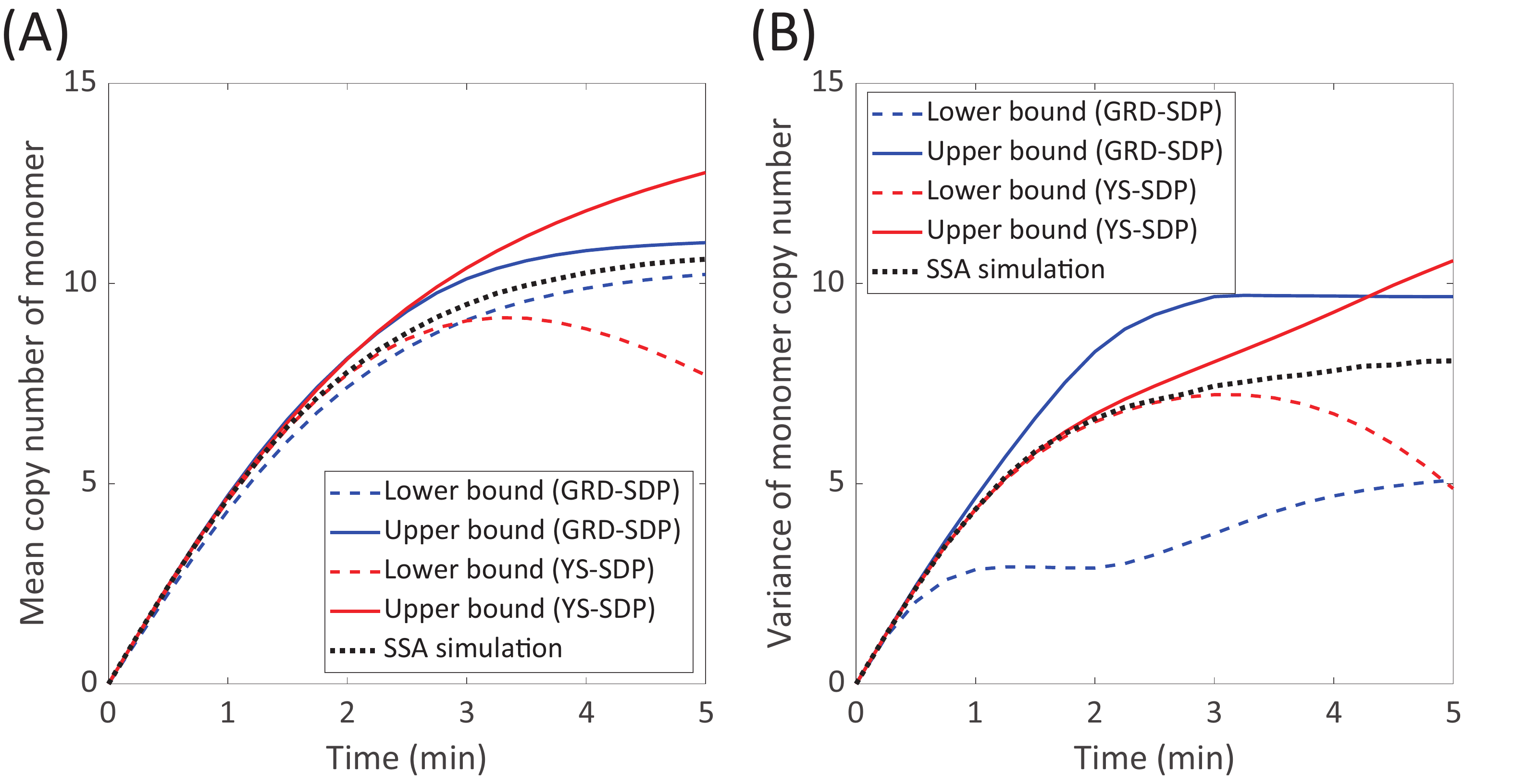}
     \caption{Bounds of the statistics of monomer protein P with YS-SDP
     and GRD-SDP. (A) the mean copy number with $\mu=7, \nu=1$ and
     $\rho_i \in \{0, -0.8844\}$ (B) the variance of the copy number
     with $\mu=7, \nu=4, \rho_i \in \{0, -0.8844, -1.7481, -2.5902, -3.4099\}$}
     \label{comp1-fig}
  \end{figure}

 \subsection{Bounds with different temporal moments}
Recently, Dowdy and Barton \cite{Dowdy2018arXiv} independently developed 
a similar optimization-based approach to obtaining the bounds of
moments. 
Therein, the following moment generating function was used instead of
the temporal moment
\begin{align}
\check{m}_{{\bm \alpha}, \rho}(0, T) := \int_{0}^{T} e^{\rho (T-t)}
 \sum_{{\bm x} \in \mathbb{K}} \prod_{j=1}^{n} x_j^{\alpha_j} P_{\bm
 x}(t) dt,
\end{align}
where $\rho \in \mathbb{R}$ is a tuning constant. 
Although the definition of $\check{m}_{{\bm \alpha}, \rho}$ loses the 
apparent connection with the necessary and sufficient moment condition
unlike ${\bm m}_{\bm \alpha, s}(t)$ (see Remark 1), 
the non-negativity of the exponential function still allows for the same  
argument that leads to a necessary condition for
the existence of a positive measure supported on
$\overline{\mathbb{K}}$ (see Proposition 1 for comparison). 
Thus, it is possible to obtain a semidefinite program of the form 
(\ref{eq:OptiProb}). %
In what follows, we use shorthands GRD-SDP and YS-SDP to refer 
to their approach and the proposed approach, respectively.

\par
\smallskip
Here we discuss how the different definitions of the moments 
affects the tightness of bounds by comparing the results of the two
optimization methods. 
Specifically, we analyzed the stochastic dimerization process in Table \ref{tbl:allwi} and a dynamic
equilibrium reaction 
\begin{align}
A + B \xrightarrow{c_1} C \xrightleftharpoons[c_3]{c_2} D \notag
\end{align}
taken from \cite{Dowdy2018arXiv}. 
To make the comparison as fair as possible, the number of decision
variables was set equal to each other. To be more specific, the dimensions of
the vectors ${\bm m}$ and ${\bm u}$ were set equal between both methods.
This means that we used the same value of $\mu$, the highest order of
moments in ${\bm m}$. 
The number of temporal moments was also set equal, that is,
$\hat{m}_{{\bm \alpha}, 0}, \hat{m}_{{\bm \alpha}, 1}, \cdots,
\hat{m}_{{\bm \alpha}, \nu}$ was used for YS-SDP and 
$\check{m}_{{\bm \alpha}, \rho_0}, \check{m}_{{\bm \alpha}, \rho_1},
\cdots,\check{m}_{{\bm \alpha}, \rho_\nu}$ for GRD-SDP.
For GRD-SDP, the constants $\rho_i$ were determined as described in \cite{Dowdy2018arXiv}. 
As a result, we obtained the same number of equality constraints
corresponding to (\ref{eq:TimeMomentEquation}). 

\par
\smallskip
Fig. \ref{comp1-fig} illustrates the bounds of the mean and the variance
of $x$ for the dimerization process. 
We observed that initially YS-SDP (proposed) computes tighter bounds 
around $t=0$, but the bounds tend to be loose as the reaction approaches
to the steady state, at which point GRD-SDP gives better bounds 
(Fig. \ref{comp1-fig}(A), (B)).
This trend hold for different choices of $\mu$ and $\nu$ (Fig. S2). 
These observations suggest that the difference of the
temporal moments may affect the frequency band of the dynamic moments at
which the bounds are tight, though the rationale needs further study in
future. 
Specifically, YS-SDP tends to give tighter bounds when the system
evolves at relatively high frequency (at the beginning of the reaction)
compared with GRD-SDP.

    \begin{figure}[tb]
   \centering
  \includegraphics[clip, width=8.5cm]{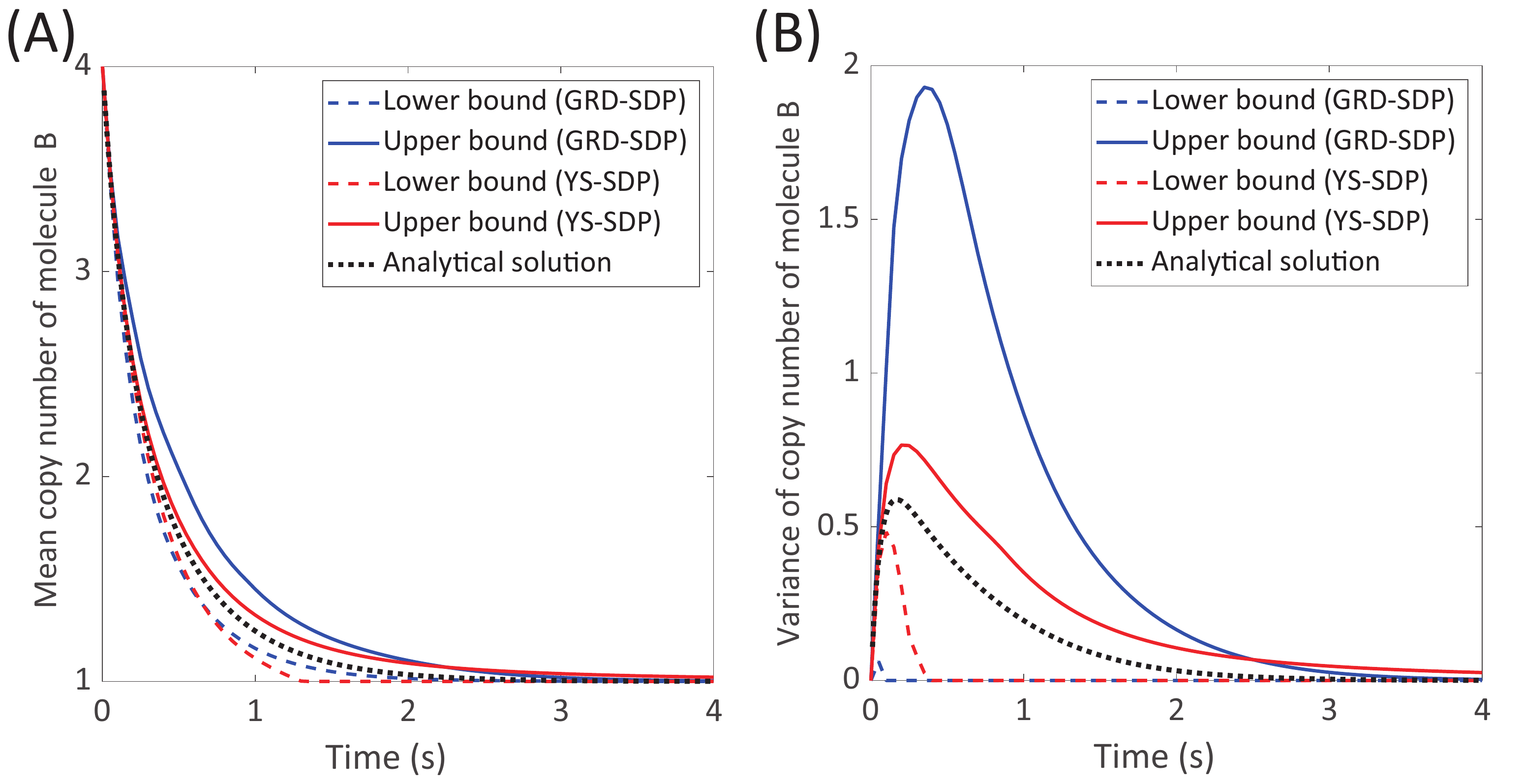}
     \caption{Bounds of the statistics of B with YS-SDP
     and GRD-SDP. (A) the mean copy number (B) the variance of the copy
     number. $\mu=3$, $\nu=2$ and $\rho_i \in \{0, -2, -2.4\}$
     were used to obtain both figures.}
     \label{comp2-fig}
  \end{figure}

\par
\smallskip
We observed the same trend for most of the reaction examples presented
in \cite{Dowdy2018arXiv} (Fig. S3-S7). The only exception was the reaction system
that involves only unimolecular reactions, that is, the cases where $w_i({\bm x})$ is
linear in ${\bm x}$, in which case GRD-SDP computed tight bounds as
explained in \cite{Dowdy2018arXiv} (Fig. S5). 
As an example, we show the results of the dynamic equilibrium reaction
in Fig. \ref{comp2-fig}(A), (B). 
In this figure, we deliberately picked $\mu$ and $\nu$ to be small so that the difference between YS-SDP and
GRD-SDP becomes clear, but we also confirmed that the bounds tend to be
tighter by making these parameters large (Fig. S8).

  \section{Conclusion}
This paper has proposed an optimization algorithm for
computing the transient statistics of stochastic chemical reactions. We
have first introduced the concept of temporal moments. 
This has enabled the derivation of an equality constraint that the transient
moments must satisfy. Although this equation is underdetermined, the possible solutions can be drastically narrowed by 
employing the moment conditions.
Combining these equality and semidefinite conditions, we have obtained
the SDP for computing the bounds of transient moments. Finally, the proposed optimization has been compared with a similar
formulation proposed in \cite{Dowdy2018arXiv} to discuss the advantages and limitations.

\bibliographystyle{ieeetr}
\bibliography{arXiv_YS_YH}

\appendix
The size of the moment matrices in the optimization problem is defined
as follows. 
For $H_0(\setm)$,
\begin{align}
\gamma_1=&\left\{\begin{array}{cl}
(\mu-1)/2&({\rm if}~\mu{\rm~is~odd})\\
\mu/2&({\rm if}~\mu{\rm~is~even})
\end{array}
 \right..
 \label{eq:gammaH0}
\end{align}
For $H(\setm, g_k)$,
\begin{align}
\gamma_1=&\left\{\begin{array}{cl}
(\mu-1)/2&({\rm if}~\mu{\rm~is~odd})\\
\mu/2 - 1&({\rm if}~\mu{\rm~is~even})
\end{array}
\right..
\end{align}
For $\hat{H}_0(\sethm, \hat{g}_k)$,
\begin{align}
\gamma_1=&\left\{\begin{array}{cl}
(\mu+1)/2&({\rm if}~\mu{\rm~is~odd})\\
\mu/2+1 & ({\rm if}~\mu~{\rm is~even~and}~\nu~{\rm is~odd})\\
\mu/2 & ({\rm if}~\mu~{\rm is~even~and}~\nu~{\rm is~even})
\end{array}
\right.\label{eq:gamma11}.
\end{align}
For $\hat{H}(\sethm, \hat{g}_k)$ with $k=1,2,\cdots,n$, $\gamma_1$ 
is defined by (\ref{eq:gammaH0}). %
For $\hat{H}(\sethm, \hat{g}_k)$ with $k=n+1, n+2$, $\gamma_1$ 
is defined by (\ref{eq:gamma11}).
For $\hat{H}_0(\sethm, \hat{g}_k)$, $\gamma_2$ is defined by
\begin{align}
\gamma_2=&\left\{\begin{array}{cl}
(\nu-1)/2&({\rm if}~\nu~{\rm~is~odd})\\
\nu/2&({\rm if}~\nu~{\rm~is~even})
\end{array}
\right.\label{eq:gamma2}.
\end{align}
For $\hat{H}(\sethm, \hat{g}_k)$ with $k=1,2,\cdots,n$, $\gamma_2$ is
 defined by (\ref{eq:gamma2}).
For $\hat{H}(\sethm, \hat{g}_k)$ with $k=n+1,n+2$,
\begin{align}
\gamma_2=&\left\{\begin{array}{cl}
(\nu-1)/2&({\rm if}~\nu~{\rm~is~odd})\\
\nu/2-1&({\rm if}~\nu~{\rm~is~even})
\end{array}
\right..
\end{align}
\end{document}